\documentclass[pra,twocolumn,shownopacs,amssymb]{revtex4}
\usepackage{graphicx,psfrag,amsmath,amssymb}
    
\begin{document}

\title{Atom-dimer and dimer-dimer scatterings in a spin-orbit coupled Fermi gas}

\author{M. Iskin}
\affiliation{Department of Physics, Ko\c{c} University, Rumelifeneri Yolu, 
34450 Sar\i yer, Istanbul, Turkey}

\date{\today}

\begin{abstract}

Using the diagrammatic approach, here we study how spin-orbit coupling (SOC) 
affects the fermion-dimer and dimer-dimer scattering lengths in the Born approximation, 
and benchmark their accuracy with the higher-order approximations. 
We consider both isotropic and Rashba couplings in three dimensions, and show 
that the Born approximation gives accurate results in the $1/(m \alpha a_s) \ll -1$ limit, 
where $m$ is the mass of the fermions, $\alpha$ is the strength of the SOC, 
and $a_s$ is the $s$-wave scattering length between fermions. This is because 
the higher-loop contributions form a perturbative series in the $1/(m \alpha a_s) < 0$ 
region that is controlled by the smallness of the residue $Z$ of the dimer propagator. 
In sharp contrast, since $Z$ grows with the square-root of the binding energy of the 
dimer in the $1/(m \alpha a_s) > 0$ region, all of the higher-loop contributions are 
of similar order.

\end{abstract}

\maketitle

\section{Introduction}
\label{sec:intro}

The diagrammatic approach has proven to be a powerful technique for studying 
few-body problems in many branches of theoretical physics. For instance, 
in the context of short-range two-body interactions between particles, it has 
been successfully applied to both the three-body~\cite{bedaque98, brodsky06, 
levinsen06, iskin08, iskin10, alzetto10, levinsen11} and four-body~\cite{pieri00, brodsky06, 
levinsen06, alzetto13} problems to verify the known exact results for the 
fermion-dimer~\cite{skorniakov57, petrov03} and dimer-dimer~\cite{petrov05} 
scattering lengths, respectively. In addition, the approach have recently 
been generalized to the three-body problem with arbitrary-range two-body 
interactions, and applied to the electron-exciton scattering in semiconductors,
i.e., to the so-called three-body Coulomb problem~\cite{combescot17}. 

Furthermore, in the context of BCS-BEC crossover~\cite{strinati18}, the 
fermion-dimer and dimer-dimer scattering lengths appear in some of the 
many-body properties of dilute Fermi gases, including their low-energy collective 
modes, superfluid density, etc.. Such appearances are quite natural in 
those parameter regimes where a strongly interacting Fermi-Fermi mixture 
can be mapped to a weakly-interacting Bose-Fermi mixture of paired 
(i.e., bosonic dimers) and unpaired (i.e., excess) fermions~\cite{pieri06, taylor07, iskin08}. 
However, it is also known that the usual treatment of the BCS-BEC crossover 
through a Gaussian fluctuation approach yields fermion-dimer and 
dimer-dimer scattering lengths that are consistent with the lowest-order 
Born approximation~\cite{brodsky06, levinsen06, pieri00}.

Given the recent surge of experimental~\cite{cheuk12, williams13, 
huang16, meng16} and theoretical~\cite{zhai11, iskin11, hu11, he12a, 
he12b, shenoy12a, shenoy12b} interests in spin-orbit-coupled Fermi gases, 
here we extend the diagrammatic approach to the relevant few-body problems.
In particular, we study how SOC affects the fermion-dimer and dimer-dimer 
scattering lengths in the Born approximation, and benchmark their accuracy 
with the higher-order approximations. Our primary findings for the isotropic and 
Rashba couplings in three dimensions are as follows. We show that the Born 
approximation gives accurate results in the $1/(m \alpha a_s) \ll -1$ limit, 
as the higher-loop contributions form a perturbative series in the 
$1/(m \alpha a_s) < 0$ region that is controlled by the residue $Z$ of the 
dimer propagator. While $Z$ decays to $0$ in the $1/(m \alpha a_s) \to -\infty$ 
limit, it grows with the square-root of the binding energy of the dimer in the 
$1/(m \alpha a_s) > 0$ region, suggesting that it may be sufficient to 
consider a finite number of higher-loop diagrams in the $1/(m \alpha a_s) < 0$ region. 

The rest of the paper is organized as follows. In Sec.~\ref{sec:obp}, we introduce 
the one-body Hamiltonian, helicity bands and the fermion propagator. 
In Sec.~\ref{sec:twbp}, we introduce the two-body Hamiltonian, identify the 
appropriate Feynman rules for the bound-state problem, and derive the 
dimer propagator for the composite bosons. 
In Sec.~\ref{sec:thbp}, we analyze the fermion-dimer scattering t-matrix, 
and extract the fermion-dimer scattering length in the zero-loop Born, 
one-loop and two-loop approximations. 
In Sec.~\ref{sec:fbp}, we analyze the dimer-dimer scattering 
t-matrix, and extract the dimer-dimer scattering length in the one-loop Born 
and two-loop approximations. In Sec.~\ref{sec:mbp}, we discuss how the
fermion-dimer and dimer-dimer scattering lengths are related to the many-body 
problem. In Sec.~\ref{sec:Rsoc}, we compare our findings for the isotropic SOC 
with those of the anisotropic (Rashba) SOC. The paper ends with a brief summary
of our conclusions in Sec.~\ref{sec:conc}. For the sake of completeness, the 
binding energy and effective mass of the dimer are presented in the Appendix.

\section{One-body problem}
\label{sec:obp}

\begin{figure} [htb]
\centerline{\scalebox{0.4}{\includegraphics{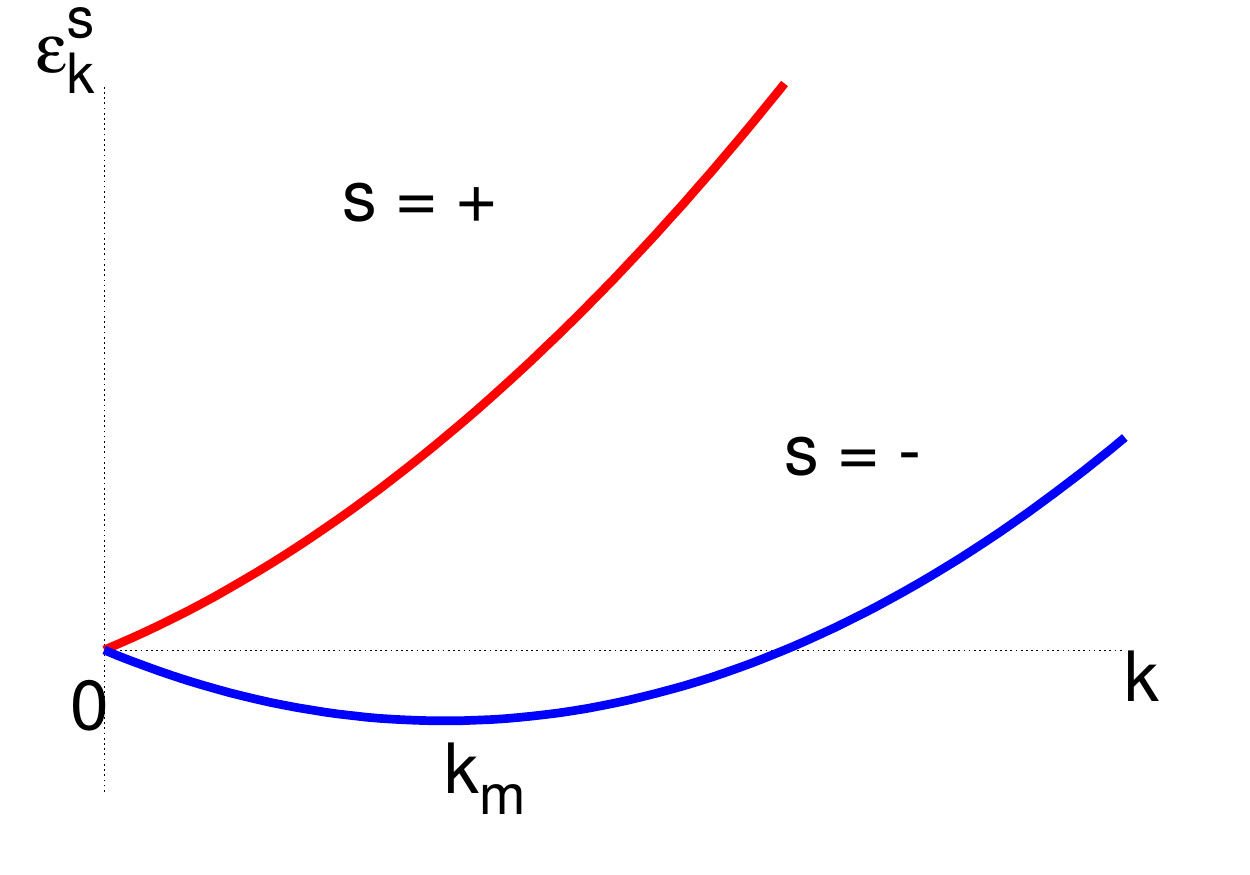}}}
\caption{\label{fig:onebody} 
One-body dispersions $\varepsilon_\mathbf{k}^s = k^2/(2m) + s \alpha k$ 
for the $s = \pm$ helicity bands. The minimum of the lower band corresponds 
to a shell of $\mathbf{k}$ states with the radius $k_m = m\alpha$ and 
energy $\varepsilon_{\mathbf{k}_m}^- = -m\alpha^2/2$.
}
\end{figure}

In the 
$ 
\langle \uparrow | = \begin{pmatrix} 1 & 0 \end{pmatrix}
$
and
$ 
\langle \downarrow | = \begin{pmatrix} 0 & 1 \end{pmatrix}
$
basis of the $\sigma_z$ Pauli matrix, the single-particle problem is governed 
by the Hamiltonian matrix
\begin{align}
\label{eqn:hk}
h_\mathbf{k} = \varepsilon_\mathbf{k} \sigma_0 + \alpha \mathbf{k} \cdot \boldsymbol{\sigma}
\end{align}
in momentum space, where $\mathbf{k} = (k_x, k_y, k_z)$ is the wave vector, 
$
\varepsilon_\mathbf{k} = k^2/(2m)
$
is the usual dispersion with $k = \sqrt{k_x^2+k_y^2+k_z^2}$ in units of $\hbar = 1$, 
$\sigma_0$ is a unit matrix, $\alpha \ge 0$ is the strength of the SOC that is taken 
as an isotropic field in $\mathbf{k}$ space, and 
$
\boldsymbol{\sigma} = (\sigma_x, \sigma_y, \sigma_z)
$ 
is a vector of Pauli matrices. The eigenvalues and eigenvectors of $h_\mathbf{k}$ 
are determined by the unitary transformation 
\begin{align}
\label{eqn:Uk}
U_\mathbf{k} = \frac{1}{\sqrt{2k(k-k_z)}}
\begin{pmatrix}
k_x - i k_y & k_z - k \\
k - k_z & k_x + i k_y
\end{pmatrix},
\end{align}
where $U_\mathbf{k}^\dagger h_\mathbf{k} U_\mathbf{k}$ gives the dispersion relations
of the $s = \pm$ helicity bands
\begin{align}
\label{eqn:esk}
\varepsilon_\mathbf{k}^s = \frac{k^2}{2m} + s \alpha k,
\end{align}
and $U_\mathbf{k} |\uparrow (\downarrow) \rangle$ gives the corresponding eigenstates. 
We illustrate these dispersions in Fig.~\ref{fig:onebody} as a function of $k$, and note 
that the ground state of the $-$-helicity band corresponds to a degenerate shell of 
$\mathbf{k}$ states with the radius $k_m = m\alpha$ and energy 
$\varepsilon_{\mathbf{k}_m}^- = -m\alpha^2/2$.

Given the Hamiltonian matrix in Eq.~(\ref{eqn:hk}), the propagator of the single
particle can be written as
\begin{align}
\label{eqn:Gk}
G(\mathbf{k}, k_0) = \frac{1}{(k_0 + i 0^+) \sigma_0 - h_\mathbf{k}},
\end{align}
where $k_0$ is the energy, and we set the chemical potential $\mu$ to $0$ 
for the few-body problems of interest below. In our analysis, we reexpress such 
propagators via the generic relation
$1/(A\sigma_0 - \mathbf{B} \cdot \boldsymbol{\sigma})
= (A\sigma_0 + \mathbf{B} \cdot \boldsymbol{\sigma})/(A^2- B^2)
= (1/2) \sum_s (\sigma_0 + s\widehat{\mathbf{B}} \cdot \boldsymbol{\sigma})/(A-sB),
$
where $\widehat{\mathbf{B}} = \mathbf{B}/B$ and $B = |\mathbf{B}|$.

\section{Two-body problem}
\label{sec:twbp}

Having in mind the atomic Fermi gases where the bosonic dimer is a result 
of a short-range interaction between $\uparrow$ and $\downarrow$ fermions, 
our two-body interaction is governed by the Hamiltonian density 
\begin{align}
\label{eqn:hr}
h_\mathbf{r} =  -g \psi_\uparrow^\dagger(\mathbf{r}) \psi_\downarrow^\dagger(\mathbf{r})
\psi_\downarrow(\mathbf{r}) \psi_\uparrow(\mathbf{r})
\end{align}
in real space, where $g \ge 0$ is the strength of the fermion-fermion attraction, 
and $\psi_\sigma^\dagger (\mathbf{r})$ and $\psi_\sigma (\mathbf{r})$ are the 
fermionic field operators. A convenient way to understand the action of this term 
is through a Hubbard-Stratonovich transformation in the imaginary-time functional 
path-integral formalism~\cite{iskin11, hu11, he12a, he12b, shenoy12a, shenoy12b}. 
Introducing the Hubbard-Stratonovich fields
$
\Delta = - g \psi_\downarrow \psi_\uparrow
$ 
and 
$
\bar{\Delta} = - g \bar{\psi}_\uparrow \bar{\psi}_\downarrow,
$ 
where $\bar{\psi}_\sigma$and $\psi_\sigma$ are the corresponding Grassmann 
variables with suppressed arguments $x = (\mathbf{r}, \tau)$  for notational 
simplicity, the action that corresponds to Eq.~(\ref{eqn:hr}) is replaced by three terms
$
\bar{\Delta} \Delta/g 
+ \Delta \bar{\psi}_\uparrow \bar{\psi}_\downarrow
+ \bar{\Delta} \psi_\downarrow \psi_\uparrow.
$
If one interprets $\bar{\Delta} = \Delta^*$ as the complex dimer field then 
the first term describes free dimers with a bare propagator $-g$, and the 
second and third terms describe the dimer-fermion and fermion-dimer 
conversion processes, respectively.

\begin{figure} [htb]
\centerline{\scalebox{1}{\includegraphics{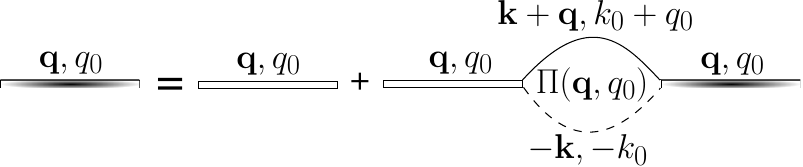}}}
\caption{\label{fig:dimer} 
Diagrammatic representation of the two-body binding problem. The dimer 
propagator (colored bars) is determined by dressing its bare propagator 
(uncolored bars) with infinitely-many fermion-fermion bubbles (solid and 
dashed lines), forming eventually a geometric series.
}
\end{figure}

The diagrammatic representation of the two-body binding problem in 
$\mathbf{k}$ space is shown in Fig.~\ref{fig:dimer}, where the physical 
dimer propagator $D(\mathbf{q}, q_0)$ is determined by dressing its bare value, 
which is a constant in space, with repeated interactions between its 
fermionic constituents~\cite{brodsky06, levinsen06}.
The resultant geometric series can be summed over to yield
\begin{align}
\label{eqn:Dqfull}
D(\mathbf{q}, q_0) = \frac{-g} {1 + g \Pi(\mathbf{q}, q_0)},
\end{align}
where $\Pi(\mathbf{q}, q_0)$ corresponds to the fermion-fermion bubble diagram
that is given by
\begin{align}
\label{eqn:Piq}
\Pi(\mathbf{q}, q_0) = \frac{\mathrm{Tr}}{2} \sum_k \sigma_y 
G(\mathbf{k}+\mathbf{q}, k_0+q_0) \sigma_y G^\mathrm{T}(-\mathbf{k}, -k_0).
\end{align}
Here, $\mathrm{Tr}$ is a trace over the spin sector, and $\sum_k$ represents 
$
\sum_{\mathbf{k}, k_0} = i \int d^3\mathbf{k} \int dk_0/(2\pi)^4.
$
In our diagrams, while the solid lines correspond to the fermion propagators 
that are described by Eq.~(\ref{eqn:Gk}), 
the dashed lines correspond to their dimer partners that are described by 
the transpose $\mathrm{T}$ of Eq.~(\ref{eqn:Gk}). This is because the dimer is 
formed between a particle that is governed by $h_{\mathbf{k}+\mathbf{q}}$ and 
a hole that is governed by $-h_{-\mathbf{k}}^\mathrm{T}$ in $\mathbf{k}$ 
space~\cite{iskin11, hu11, he12a, he12b, shenoy12a, shenoy12b}.
In accordance with the Feynman rules, each fermion line, dimer line and vertex 
carries a factor of $i$. In addition, we associate each dimer-creation 
(-annihilation) vertex with an additional factor of $\mp i \sigma_y$ to 
account for the fermion-dimer (dimer-fermion) conversion terms, i.e., 
$-i \bar{\Delta} \sigma_y$ and $i \Delta \sigma_y$, respectively, in the 
particle-hole sectors. 

Noting the relation
$
\sigma_y \mathbf{B} \cdot \boldsymbol{\sigma}^\mathrm{T} \sigma_y 
= - \mathbf{B} \cdot \boldsymbol{\sigma},
$
and integrating $k_0$ in the upper half plane in which there are two simple 
poles at $k_0 = - \varepsilon_\mathbf{k}^\pm$, we find~\cite{he12b} 
\begin{align}
\label{eqn:Piq0}
\Pi(\mathbf{q}, q_0) = \frac{1}{4} \sum_{ss' \mathbf{k}} 
\frac{1 + ss' \widehat{\mathbf{k}} \cdot \widehat{\mathbf{Q}}}
{q_0 - \varepsilon_\mathbf{k}^s - \varepsilon_\mathbf{Q}^{s'}},
\end{align}
where $\mathbf{Q} = \mathbf{k}+\mathbf{q}$. Equation~(\ref{eqn:Piq0}) shows that 
only the intra-band processes contribute to the bubble diagram when the dimer 
is stationary, i.e., when its center-of-mass momentum $\mathbf{q}$ vanishes. 
Therefore, we can reexpress the stationary bubble diagram as
$
\Pi(\mathbf{0}, q_0) = (1/2) \sum_{s k} G_s(\mathbf{k}, k_0+q_0) G_s(-\mathbf{k}, -k_0),
$
where 
\begin{align}
\label{eqn:Gsk}
G_s(\mathbf{k}, k_0) = \frac{1} {k_0 - \varepsilon_\mathbf{k}^s + i0^+}
\end{align}
is the fermion propagator in the $s = \pm$ helicity basis.

In the lowest order in $\mathbf{q}$ and $q_0$, Eq.~(\ref{eqn:Dqfull}) has the 
generic structure of a simple pole~\cite{iskin11, hu11, he12a, he12b, shenoy12a, shenoy12b} 
\begin{align}
\label{eqn:Dq}
D(\mathbf{q}, q_0) = \frac{Z} {q_0 - q^2/(2m_B) + \mu_B + i0^+}
\end{align}
where 
$
Z = 8\pi(|\varepsilon_b| - m\alpha^2)^{3/2}/(m\sqrt{m} |\varepsilon_b|)
$
corresponds to the residue of the pole,
$
2m/m_B = 7/3 - 4(1 - m\alpha^2/|\varepsilon_b|)^{3/2}/3 - 2m\alpha^2/|\varepsilon_b|
$
determines the effective mass of the bosonic dimer, and
$\mu_B = 2\mu - \varepsilon_b \to -\varepsilon_b$ corresponds to its chemical potential.
Noting that $-m\alpha^2$ is the two-body continuum threshold, the energy of the 
two-body bound state $\varepsilon_b \le -m\alpha^2$ or the two-body binding energy 
$|\varepsilon_b|-m\alpha^2$ of the dimer can be simply found by looking at the pole 
of $D(\mathbf{0}, \varepsilon_b)$, leading to the relation 
$
1 = (g/2) \sum_{s \mathbf{k}} 1/(2\varepsilon_\mathbf{k}^s - \varepsilon_b).
$
In addition, we substitute $g$ with the usual t-matrix relation between two fermions in 
vacuum without the SOC, 
$
1/g = -m V/(4\pi a_s) + \sum_\mathbf{k} 1/(2\varepsilon_\mathbf{k}),
$ 
where $\sum_\mathbf{k} = V \int d^3\mathbf{k}/(2\pi)^3$ in units of $V = 1$. 
This leads to
$
\varepsilon_b = -2m \alpha^2 - 1/(2ma_s^2) \pm \sqrt{1/(4m^2a_s^4) + \alpha^2/a_s^2}
$
for $a_s \lessgtr 0$, showing that $\varepsilon_b \le -m\alpha^2$ for all parameters.
This expression is analytically tractable in three limits~\cite{iskin11, he12b, shenoy12a, 
shenoy12b}: we find that
(i) $\varepsilon_b = -m\alpha^2 - m^3\alpha^4a_s^2$ and $m_B = 6m$ 
in the limit when $1/(m \alpha a_s) \ll -1$,
(ii) $\varepsilon_b = -2m\alpha^2$ and $m_B = 3\sqrt{2}/(2\sqrt{2}-1) m \approx  2.32 m$
in the unitarity limit when $1/(m \alpha a_s) = 0$, and 
(iii) $\varepsilon_b = -1/(ma_s^2)$ and $m_B = 2m$ 
in the limit when $1/(m \alpha a_s) \gg 1$.
These results are illustrated in Fig.~\ref{fig:EbmB} for the completeness
of the presentation. Note that the latter limit recovers the usual two-body problem 
with no SOC in the $1/(m \alpha a_s) \gg 1$ limit when $\alpha \to 0^+$.

\section{Three-body problem}
\label{sec:thbp}

\begin{figure} [htb]
\centerline{\scalebox{1}{\includegraphics{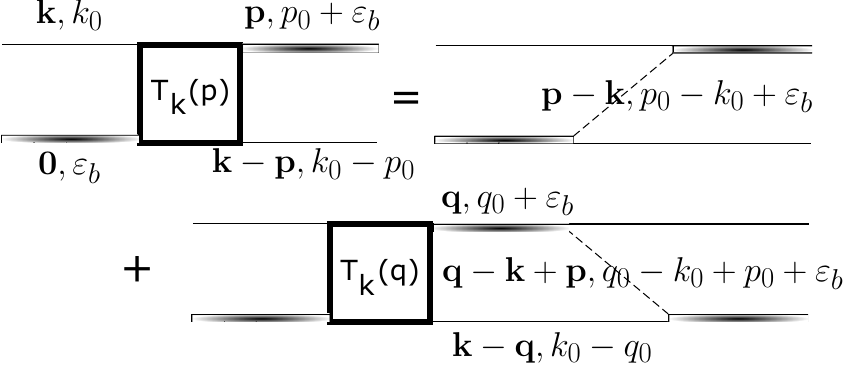}}}
\caption{\label{fig:tmatrix} 
Diagrammatic representation of the three-body problem. The fermion-dimer 
scattering t-matrix is determined by repeating the fermion-exchange process 
infinitely-many times, forming eventually an integral equation.
}
\end{figure}

In this section, we are interested in the scattering t-matrix $t_k^{--}(0)$ between the 
lowest-energy fermion in the $-$-helicity band and a stationary dimer. 
For this purpose, we introduce a shorthand notation
\begin{align}
\mathcal{T}_k(p) =
\begin{bmatrix}
t_k^{\uparrow\uparrow}(p) & t_k^{\uparrow\downarrow}(p) \\
t_k^{\downarrow\uparrow}(p) & t_k^{\downarrow\downarrow}(p)
\end{bmatrix},
\end{align}
where $k = ({\mathbf{k},k_0})$ refers collectively to the momentum and 
energy of the incoming fermion, and $p=({\mathbf{p},p_0})$ refers 
collectively to the momentum and energy exchange between the 
outgoing fermion and the dimer. We refer to Fig.~\ref{fig:tmatrix} for 
the clarity of its meaning. Once $\mathcal{T}_k(0)$ is evaluated, 
we transform it to the helicity basis via Eq.~(\ref{eqn:Uk}), and obtain
$
U_\mathbf{k}^\dagger \mathcal{T}_k(0) U_\mathbf{k}.
$
Using the spherical coordinates where
$
\widehat{\mathbf{k}} = (  \sin \theta_\mathbf{k} \cos \phi_\mathbf{k}, 
\sin \theta_\mathbf{k} \sin \phi_\mathbf{k}, \cos \theta_\mathbf{k}  ),
$
we find
\begin{align}
t_k^{ss} (0) &= t_k^{\uparrow \uparrow} (0) \frac{1 + s \cos \theta_\mathbf{k}}{2}
+ t_k^{\downarrow \downarrow} (0) \frac{1 - s \cos \theta_\mathbf{k}}{2} \nonumber \\
&+ s \mathrm{Re} [t_k^{\uparrow \downarrow} (0) \sin \theta_\mathbf{k} 
(\cos \phi_\mathbf{k} + i\sin \phi_\mathbf{k}) ]
\label{eqn:tkss}
\end{align}
for the diagonal elements with $\mathrm{Re}$ the real part. 
Note in particular that $t_k^{--} (0) = t_k^{\downarrow \downarrow} (0)$ for 
$\mathbf{k}$ that is aligned with the $z$ axis, i.e., when $\theta_\mathbf{k} = 0$.
In this paper, we are interested in the fermion-dimer scattering length 
$a_{BF}$ that is determined by~\cite{brodsky06, levinsen06}
\begin{align}
\label{eqn:abft}
a_{BF} = \frac{m_{BF}}{4\pi} Z t_k^{--} (0),
\end{align}
where $m_{BF} = 2m_B m_F / (m_B+m_F)$ is twice the reduced mass
of the fermion and the dimer, and 
$
k = (m\alpha \widehat{\mathbf{k}}, -m\alpha^2/2)
$ 
corresponds to the lowest-energy eigenstate in the $-$-helicity band.

\begin{figure} [htb]
\centerline{\scalebox{1}{\includegraphics{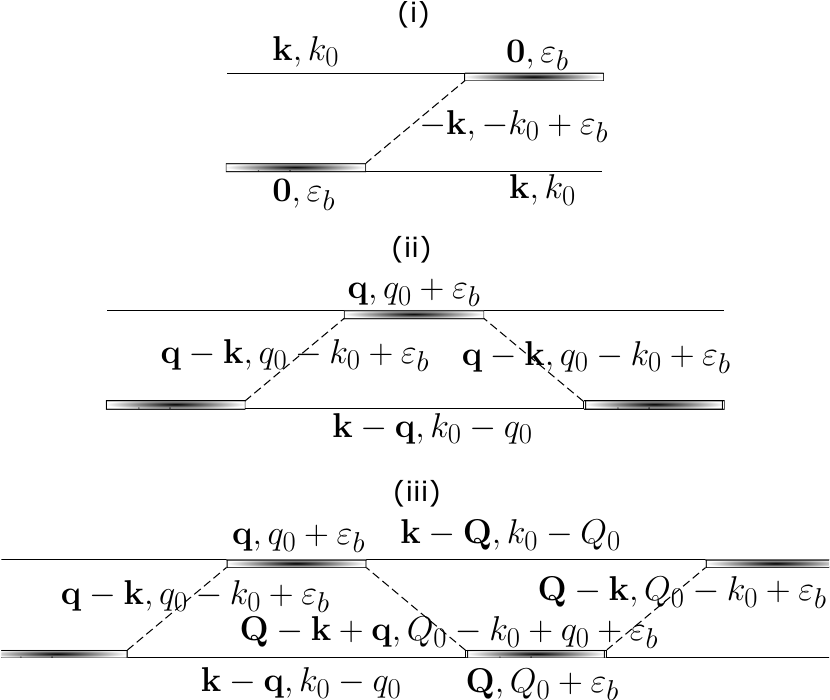}}}
\caption{\label{fig:loops} 
Diagrammatic representations of the (i) zero-loop Born, (ii) one-loop and 
(iii) two-loop contributions to the fermion-dimer scattering t-matrix. 
}
\end{figure}

\begin{figure} [htb]
\centerline{\scalebox{0.5}{\includegraphics{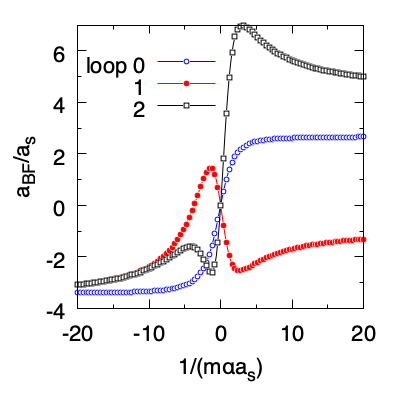}}}
\caption{\label{fig:abf} 
Fermion-dimer scattering length in the zero-loop Born, one-loop and two-loop 
approximations.
The higher-loop contributions form a perturbative series in the $1/(m \alpha a_s) < 0$
region, but they are of similar order in the $1/(m \alpha a_s) > 0$ region.
}
\end{figure}

For instance, the diagrammatic representations of the zero-loop, one-loop and 
two-loop contributions to the fermion-dimer scattering t-matrix are shown 
in Fig.~\ref{fig:loops}~\cite{bedaque98, brodsky06, levinsen06, iskin08, 
iskin10, alzetto10, levinsen11, combescot17}. 
The zero-loop contribution is known as the Born approximation, and in accordance 
with the Feynman rules given above, it is given by
\begin{align}
\label{eqn:T0}
\mathcal{T}_k^{(0)}(0) = - \sigma_y G^\mathrm{T}(-\mathbf{k},-k_0+\varepsilon_b) \sigma_y,
\end{align}
where the minus sign is due to the exchange of an identical fermion.
By plugging Eq.~(\ref{eqn:T0}) into Eq.~(\ref{eqn:tkss}), we find
\begin{align}
\label{eqn:tkssb}
{t_k^{ss}}^{(0)} (0) = \frac{1} {k_0 - \varepsilon_b + \varepsilon_\mathbf{k}^s},
\end{align}
which is physically intuitive. This is because, since both dimers are stationary in 
the Born diagram, the helicity bands are not coupled, and $t_k^{ss} (0)$ can 
be directly expressed as
$
{t_k^{ss}}^{(0)} (0) = -G_s(-\mathbf{k}, -k_0 + \varepsilon_b).
$
Furthermore, by plugging ${t_k^{--}}^{(0)} (0) = 1/(|\varepsilon_b| - m\alpha^2)$ 
into Eq.~(\ref{eqn:abft}), we find
\begin{align}
\label{eqn:abfborn}
a_{BF}^\mathrm{Born} = \frac{2 m_{BF} \sqrt{|\varepsilon_b| - m\alpha^2}}
{m \sqrt {m} |\varepsilon_b|}
\end{align}
in the Born approximation, suggesting that the fermion-dimer interaction is 
repulsive for all parameters. In Fig.~\ref{fig:abf}, we show $a_{BF}^\mathrm{Born}$ 
as a function of $1/(m \alpha a_s)$, which is analytically tractable in three limits: 
(i) $a_{BF}^\mathrm{Born} = -24a_s/7$ in the limit when 
$1/(m \alpha a_s) \ll -1$, 
(ii) $a_{BF}^\mathrm{Born} = 6\sqrt{2}/[m\alpha (5\sqrt{2}-1)] \approx  1.40/ (m \alpha)$ 
in the unitarity limit when $1/(m \alpha a_s) = 0$, and
(iii) $a_{BF}^\mathrm{Born} = 8a_s/3$ in the limit when $1/(m \alpha a_s) \gg 1$.
Note that the latter limit recovers the usual three-body problem with no SOC 
in the $1/(m \alpha a_s) \gg 1$ limit when $\alpha \to 0^+$.

To go beyond the Born approximation, we consider the one-loop contribution that 
is shown in Fig.~\ref{fig:loops}(ii). In accordance with the Feynman rules, 
this diagram is given by 
\begin{align}
&\mathcal{T}_k^{(1)} (0) = \sum_q
\sigma_y G^\mathrm{T}(\mathbf{q}-\mathbf{k}, q_0-k_0+\varepsilon_b) \sigma_y 
D(\mathbf{q}, q_0+\varepsilon_b)  \nonumber \\ 
&\times G(\mathbf{k}-\mathbf{q}, k_0-q_0) \sigma_y 
G^\mathrm{T}(\mathbf{q}-\mathbf{k}, q_0-k_0+\varepsilon_b) \sigma_y.
\end{align}
Noting the relation
$
(\sigma_0 + \widehat{\mathbf{A}} \cdot \boldsymbol{\sigma})
(\sigma_0 + \widehat{\mathbf{B}} \cdot \boldsymbol{\sigma})
= (1 + \widehat{\mathbf{A}} \cdot \widehat{\mathbf{B}}) \sigma_0
+ (\widehat{\mathbf{A}} + \widehat{\mathbf{B}} + i \widehat{\mathbf{A}} \times \widehat{\mathbf{B}})
\cdot \boldsymbol{\sigma},
$
we first integrate $q_0$ in the upper half plane in which there are two simple 
poles at $q_0 = k_0 - \varepsilon_\mathbf{q}^\pm$, and reduce the t-matrix 
contribution to
\begin{align}
\mathcal{T}_k^{(1)} (0) = \frac{1}{2}\sum_{s \mathbf{q}}
\frac { D(\mathbf{q}+\mathbf{k}, k_0 + \varepsilon_b - \varepsilon_\mathbf{q}^s) }
{ (2\varepsilon_\mathbf{q}^s - \varepsilon_b)^2 } 
(\sigma_0 - s \widehat{\mathbf{q}} \cdot \boldsymbol{\sigma}).
\end{align}
Noting that $\mathcal{T}_k(0)$ has a spherical symmetry in $\mathbf{k}$ space, 
we choose an incoming momentum $\mathbf{k} = m\alpha \widehat{k_z}$ that
is aligned with the $z$ axis, and perform the remaining integrations numerically 
in the $\mathbf{q}$ space~\cite{integralnote}. 
In Fig.~\ref{fig:abf}, we show how the one-loop contribution 
affects $a_{BF}^\mathrm{Born}$ as a function of $1/(m \alpha a_s)$. 
In the one-loop approximation, we find that $a_{BF}^\mathrm{Born}$ becomes 
attractive in the $1/(m \alpha a_s) > 0$ region, which is not physical.

To go further beyond the Born approximation, we next consider the two-loop 
contribution that is represented in Fig.~\ref{fig:loops}(iii). In accordance with the 
Feynman rules, this diagram is given by 
\begin{align}
& \mathcal{T}_k^{(2)} (0) = \sum_{qQ}
\sigma_y G^\mathrm{T}(\mathbf{q}-\mathbf{k}, q_0-k_0+\varepsilon_b) \sigma_y 
D(\mathbf{q}, q_0+\varepsilon_b)  \nonumber \\
&\times G(\mathbf{k}-\mathbf{q}, k_0-q_0) 
\sigma_y G^\mathrm{T}(\mathbf{Q}-\mathbf{k}+\mathbf{q}, Q_0-k_0+q_0+\varepsilon_b) \nonumber\\
&\times \sigma_y G(\mathbf{k}-\mathbf{Q}, k_0-Q_0) D(\mathbf{Q}, Q_0+\varepsilon_b)  \sigma_y \nonumber\\
&\times G^\mathrm{T}(\mathbf{Q}-\mathbf{k}, Q_0-k_0+\varepsilon_b) \sigma_y,
\end{align}
where a minus sign is included due to the fermion exchange.
We integrate $q_0$ and $Q_0$ in their upper half planes in which there are two
simple poles at $q_0 = k_0 - \varepsilon_{\mathbf{q}-\mathbf{k}}^\pm$ and two 
simple poles at $Q_0 = k_0 - \varepsilon_{\mathbf{Q}-\mathbf{k}}^\pm$. 
In addition, by taking advantage of the symmetry of the diagram with respect to 
the internal variables $\mathbf{q}$ and $\mathbf{Q}$, we reduce the t-matrix 
contribution to
\begin{widetext}
\begin{align}
\mathcal{T}_k^{(2)} (0) = \frac{1}{8} \sum_{ss's'' \mathbf{q} \mathbf{Q}} 
\frac { D(\mathbf{q}+\mathbf{k}, k_0 + \varepsilon_b - \varepsilon_\mathbf{q}^s)
D(\mathbf{Q}+\mathbf{k}, k_0 + \varepsilon_b - \varepsilon_\mathbf{Q}^{s'}) }
{(2\varepsilon_\mathbf{q}^s - \varepsilon_b) 
(2\varepsilon_\mathbf{Q}^{s'} - \varepsilon_b) 
(\varepsilon_\mathbf{q}^s + \varepsilon_\mathbf{Q}^{s'} + \varepsilon_\mathbf{K}^{s''} - k_0 - \varepsilon_b)} 
\bigg( \big\lbrace 1 + ss'' \widehat{\mathbf{q}} \cdot \widehat{\mathbf{K}} + s's'' \widehat{\mathbf{Q}} \cdot \widehat{\mathbf{K}} + 
ss' \widehat{\mathbf{q}} \cdot \widehat{\mathbf{Q}} \big\rbrace \sigma_0 \nonumber \\
+ 
\big\lbrace s\widehat{\mathbf{q}} + s'\widehat{\mathbf{Q}} + s''\widehat{\mathbf{K}} 
- ss's'' [(\widehat{\mathbf{q}} \cdot \widehat{\mathbf{Q}}) \widehat{\mathbf{K}} 
- (\widehat{\mathbf{Q}} \cdot \widehat{\mathbf{K}}) \widehat{\mathbf{q}}/2 -
(\widehat{\mathbf{q}} \cdot \widehat{\mathbf{K}}) \widehat{\mathbf{Q}}/2)] \big\rbrace \cdot \boldsymbol{\sigma}
\bigg),
\end{align}
\end{widetext}
where
$
\mathbf{K} = \mathbf{Q} + \mathbf{q} + \mathbf{k}.
$
We again choose an incoming momentum $\mathbf{k} = m\alpha \widehat{k_z}$ 
that is aligned with the $z$ axis, and perform the remaining integrations numerically 
in the $\mathbf{q}$ and $\mathbf{Q}$ spaces~\cite{integralnote}. 
In Fig.~\ref{fig:abf}, we show how the 
combination of the one-loop and two-loop contributions affects $a_{BF}^\mathrm{Born}$ 
as a function of $1/(m \alpha a_s)$. While the two-loop contribution is negligible 
in the $1/(m \alpha a_s) \ll -1$ limit, it leads to a repulsive $a_{BF}$ in the 
$1/(m \alpha a_s) > 0$ region. 

By comparing the zero-loop, one-loop and two-loop approximations in Fig.~\ref{fig:abf}, 
we observe that while the higher-loop contributions form a perturbative series 
in the $1/(m \alpha a_s) < 0$ region, they are of similar order in the 
$1/(m \alpha a_s) > 0$ region. Noting that $Z$ decays to $0$ in the 
$1/(m \alpha a_s) \to -\infty$ limit, and that it increases as $\sqrt{|\varepsilon_b|}$ 
in the $1/(m \alpha a_s) \to +\infty$ limit, this observation is caused by the 
incremental growth of the power of $Z$ that is coming from the additional 
dimer propagators within each loop. For this reason, a proper description of the 
latter region requires infinitely-many loop diagrams at all 
orders~\cite{bedaque98, brodsky06, levinsen06, iskin08, iskin10, alzetto10, levinsen11, combescot17}. 
A practical way to handle such summations is presented in Fig.~\ref{fig:tmatrix}, 
where the fermion-dimer scattering t-matrix is determined by repeating the 
fermion-exchange process infinitely-many times, forming eventually an integral 
equation. In accordance with the Feynman rules, this diagram is given by
\begin{align}
\mathcal{T}_k(p) &=
- \sigma_y G^\mathrm{T}(\mathbf{p}-\mathbf{k}, p_0-k_0+\varepsilon_b) \sigma_y \nonumber \\
&- \sum_q 
\mathcal{T}_k(q) D(\mathbf{q}, q_0+\varepsilon_b) G(\mathbf{k}-\mathbf{q}, k_0-q_0) \nonumber \\
&\times \sigma_y G^\mathrm{T}(\mathbf{q}-\mathbf{k}+\mathbf{p}, q_0-k_0+p_0+\varepsilon_b) \sigma_y,
\end{align}
where the minus signs are due to the fermion exchanges.
Integrating $q_0$ in the upper half plane where $\mathcal{T}_k(q)$ is analytic
and there are two simple poles at $q_0 = k_0 - \varepsilon_{\mathbf{q}-\mathbf{k}}^\pm$, 
we reduce the t-matrix equation to
\begin{align}
\label{eqn:tmatrix}
&\mathcal{T}_k(\mathbf{p}, p_0) = - \frac{1}{2} \sum_s \frac
{\sigma_0 - s \widehat{\mathbf{k'}} \cdot \boldsymbol{\sigma} }
{ p_0-k_0+\varepsilon_b-\varepsilon_\mathbf{k'}^s } \\
&- \frac{1}{4} \sum_{s s' \mathbf{q}} 
\frac{ D(\mathbf{q}, k_0 + \varepsilon_b - \varepsilon_\mathbf{Q}^s) }
{p_0 + \varepsilon_b - \varepsilon_\mathbf{Q}^s - \varepsilon_\mathbf{K}^{s'} }
\mathcal{T}_k(\mathbf{q}, k_0 - \varepsilon_\mathbf{Q}^s) \nonumber \\ 
& \times [(1-ss' \widehat{\mathbf{Q}} \cdot \widehat{\mathbf{K}})\sigma_0 + 
(s \widehat{\mathbf{Q}} - s' \widehat{\mathbf{K}} - i s s' \widehat{\mathbf{Q}} \times \widehat{\mathbf{K}}) \cdot \boldsymbol{\sigma}] \nonumber.
\end{align}
Here
$\mathbf{k'} = \mathbf{p} - \mathbf{k}$,
$\mathbf{Q} = \mathbf{k} - \mathbf{q}$ and
$\mathbf{K} = \mathbf{p} - \mathbf{k} + \mathbf{q}$ are introduced for the simplicity 
of the presentation. 

In the usual three-body problem with no SOC, $t_0(\mathbf{p}, p_0)$ is not 
only a real function but it is also restricted to the so-called on-the-shell value 
$
t_0[\mathbf{p}, p_0 = -p^2/(2m)]
$
for both the incoming and outgoing fermions~\cite{bedaque98, brodsky06, 
levinsen06, iskin08, iskin10, alzetto10, levinsen11, combescot17}. In addition, 
using the spherical symmetry of the t-matrix, the problem reduces to a simple integral 
equation with a single variable for $t_0(|\mathbf{p}|)$, whose numerical computation
converges very fast. 
However, since the helicity bands are coupled due to the non-stationary dimers, there 
are two shells contributing to Eq.~(\ref{eqn:tmatrix}). Furthermore, given that the 
t-matrix is a $2 \times 2$ matrix with complex elements, this reduces Eq.~(\ref{eqn:tmatrix}) 
to an eight coupled integral equations. Unfortunately, this is quite complicated, and the 
exact numerical solution of the three-body problem remains an open problem.

\section{Four-body problem}
\label{sec:fbp}

Motivated by the overall success of the Born approximation in the fermion-dimer 
scattering problem, here we apply the diagrammatic approach to the 
scattering t-matrix $t_0^{BB}(0)$ between two stationary dimers in the one-loop 
Born and two-loop approximations. Despite its simplicity, we expect 
$a_{BB}^\mathrm{Born}$ to be quite accurate in the $1/(m \alpha a_s) \ll -1$ limit 
as the higher-order contributions form a perturbative series in the $1/(m \alpha a_s) < 0$
region. However, our results are only qualitative in the $1/(m \alpha a_s) > 0$ region, 
whose accurate description is beyond the scope of this paper~\cite{brodsky06, 
levinsen06, alzetto13} .

\begin{figure} [htb]
\centerline{\scalebox{1}{\includegraphics{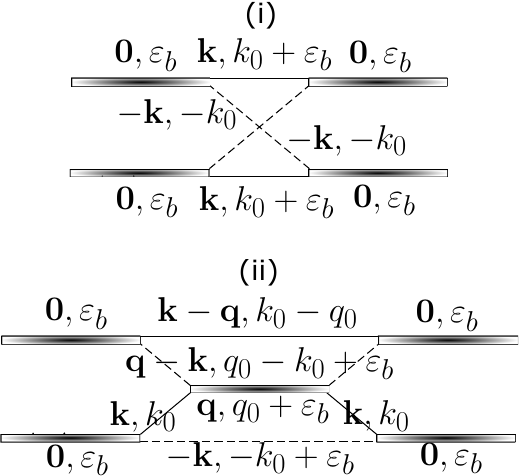}}}
\caption{\label{fig:4body} 
Diagrammatic representation of the (i) one-loop Born and (ii) two-loop contributions to 
the dimer-dimer scattering t-matrix. 
}
\end{figure}

The diagrammatic representation of the Born contribution to the dimer-dimer 
scattering t-matrix is shown in Fig.~\ref{fig:4body}(i)~\cite{pieri00, brodsky06, 
levinsen06, alzetto13}. In accordance with the Feynman rules, it is given by
\begin{align}
{t_0^{BB}}^{(1)} (0) = -\frac{\mathrm{Tr}}{2} \sum_k 
\big[ \sigma_y G(\mathbf{k}, k_0+\varepsilon_b) \sigma_y G^\mathrm{T}(-\mathbf{k}, -k_0) \big]^2,
\end{align}
where the minus sign is due to the fermion exchange.
Noting the relations
$
(\sigma_0 \pm \widehat{\mathbf{A}} \cdot \boldsymbol{\sigma})
(\sigma_0 \pm \widehat{\mathbf{A}} \cdot \boldsymbol{\sigma})
= 2 (\sigma_0 \pm \widehat{\mathbf{A}} \cdot \boldsymbol{\sigma}),
$
and
$
(\sigma_0 \pm \widehat{\mathbf{A}} \cdot \boldsymbol{\sigma})
(\sigma_0 \mp \widehat{\mathbf{A}} \cdot \boldsymbol{\sigma}) = 0,
$
we first integrate $k_0$ in the upper half plane in which there are two double 
poles at $k_0 = - \varepsilon_\mathbf{k}^\pm$, and reduce the t-matrix contribution to
\begin{align}
\label{eqn:tbb0}
{t_0^{BB}}^{(1)} (0) = \frac{1}{2} \sum_{s \mathbf{k}} \frac{2}{(2\varepsilon_\mathbf{k}^s - \varepsilon_b)^3}.
\end{align}
This is a physically intuitive result because, since all dimers are stationary in 
the Born diagram, the helicity bands are not coupled, and the diagram can 
be directly expressed as
$
{t_0^{BB}}^{(1)} (0) = (-1/2) \sum_{s k} \big[G_s(\mathbf{k}, k_0+\varepsilon_b) 
G_s(-\mathbf{k}, -k_0)\big]^2.
$

\begin{figure} [htb]
\centerline{\scalebox{0.5}{\includegraphics{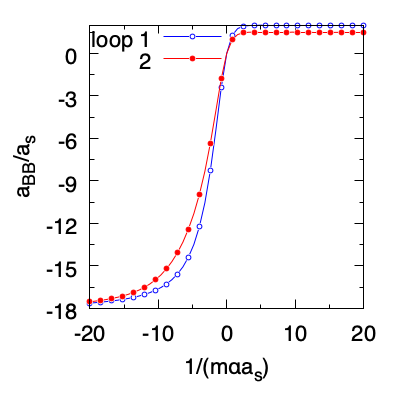}}}
\caption{\label{fig:abb} 
Dimer-dimer scattering length in the one-loop Born and two-loop approximations.
The higher-loop contributions form a perturbative series in the $1/(m \alpha a_s) < 0$
region, but they are of similar order in the $1/(m \alpha a_s) > 0$ region.
}
\end{figure}

In this paper, we are interested in the dimer-dimer scattering length $a_{BB}$ that 
is determined by~\cite{pieri00, brodsky06, levinsen06, alzetto13} 
\begin{align}
\label{eqn:abb}
a_{BB} = \frac{m_B} {4\pi} Z^2 t_0^{BB}(0) .
\end{align}
Plugging 
$
{t_0^{BB}}^{(1)} (0) = m\sqrt{m}(|\varepsilon_b| + 2m\alpha^2)/[16 \pi (|\varepsilon_b| - m\alpha^2)^{5/2}]
$ 
above, we find~\cite{he12b, shenoy12a, shenoy12b}
\begin{align}
\label{eqn:abbborn}
a_{BB}^\mathrm{Born} = \frac{m_B (|\varepsilon_b| + 2m\alpha^2) \sqrt{|\varepsilon_b| - m\alpha^2}} 
{m\sqrt{m}|\varepsilon_b|^2},
\end{align}
in the Born approximation, which suggests that the dimer-dimer interaction is 
repulsive for all parameters. In Fig.~\ref{fig:abb}, we show $a_{BB}^\mathrm{Born}$ 
as a function of $1/(m \alpha a_s)$, which is analytically tractable in three limits: 
(i) $a_{BB}^\mathrm{Born} = -18a_s$ in the limit when 
$1/(m \alpha a_s) \ll -1$, 
(ii) $a_{BB}^\mathrm{Born} = 3\sqrt{2}/[m\alpha (2\sqrt{2}-1)] \approx  2.32/ (m \alpha)$ 
in the unitarity limit when $1/(m \alpha a_s) = 0$, and
(iii) $a_{BB}^\mathrm{Born} = 2a_s$ in the limit when $1/(m \alpha a_s) \gg 1$.
Note that the latter limit recovers the usual four-body problem with no SOC 
in the $1/(m \alpha a_s) \gg 1$ limit when $\alpha \to 0^+$.

To go beyond the Born approximation, we consider the two-loop contribution that 
is shown in Fig.~\ref{fig:4body}(ii). In accordance with the Feynman rules, 
this diagram is given by 
\begin{align}
&{t_0^{BB}}^{(2)} (0) = - \frac{\mathrm{Tr}}{2} \sum_{k q}
\sigma_y G^\mathrm{T}(\mathbf{q}-\mathbf{k}, q_0-k_0+\varepsilon_b) \sigma_y \nonumber \\
&G(\mathbf{k}, k_0) D(\mathbf{q}, q_0+\varepsilon_b) 
\sigma_y G^\mathrm{T}(-\mathbf{k},-k_0+\varepsilon_b) \sigma_y G(\mathbf{k}, k_0) \nonumber \\
&\sigma_y G^\mathrm{T}(\mathbf{q}-\mathbf{k}, q_0-k_0+\varepsilon_b) 
\sigma_y G(\mathbf{k}-\mathbf{q}, k_0-q_0).
\end{align}
We first integrate $q_0$ in the upper half plane in which there are two simple 
poles at $q_0 = k_0 - \varepsilon_{\mathbf{q}-\mathbf{k}}^\pm$, and then 
integrate $k_0$ in the upper half plane in which there are two simple 
poles at $k_0 = \varepsilon_b - \varepsilon_\mathbf{k}^\pm$. This reduces
the t-matrix contribution to
\begin{align}
{t_0^{BB}}^{(2)} (0) = \frac{1}{4}\sum_{ss' \mathbf{k} \mathbf{q}}
\frac { D(\mathbf{q}, 2\varepsilon_b - \varepsilon_\mathbf{k}^s - \varepsilon_\mathbf{Q}^{s'}) }
{ (2\varepsilon_\mathbf{k}^s - \varepsilon_b)^2 
(2\varepsilon_\mathbf{Q}^{s'} - \varepsilon_b)^2} 
(1 + ss' \widehat{\mathbf{k}} \cdot \widehat{\mathbf{Q}}),
\end{align}
where $\mathbf{Q} = \mathbf{k} + \mathbf{q}$, and the remaining integrations 
are performed numerically in the $\mathbf{k}$ and $\mathbf{q}$ spaces~\cite{integralnote}. 
In Fig.~\ref{fig:abb}, we show how the two-loop 
contribution affects $a_{BB}^\mathrm{Born}$ as a function of $1/(m \alpha a_s)$. 
While the two-loop contribution is negligible in the $1/(m \alpha a_s) \ll -1$ limit, 
it is comparable to $a_{BB}^\mathrm{Born}$ in the $1/(m \alpha a_s) > 0$ region.

\section{Many-body problem}
\label{sec:mbp}

The fermion-dimer and dimer-dimer scattering lengths offer valuable insights 
for some of the many-body properties of Fermi gases. For instance, in the 
case of population-imbalanced Fermi gases, $a_{BF}$ and $a_{BB}$ 
can be used to map the strongly-interacting Fermi-Fermi mixture of $\uparrow$ 
and $\downarrow$ fermions to a weakly-interacting Bose-Fermi mixture 
of paired fermions (dimers) and unpaired (excess) ones~\cite{pieri06, taylor07, iskin08}. 
In the parameter regime where this effective description holds, the existing 
literature on true Bose-Fermi mixtures can be easily utilized to characterize 
the imbalanced Fermi gases. 

\begin{figure} [htb]
\centerline{\scalebox{0.5}{\includegraphics{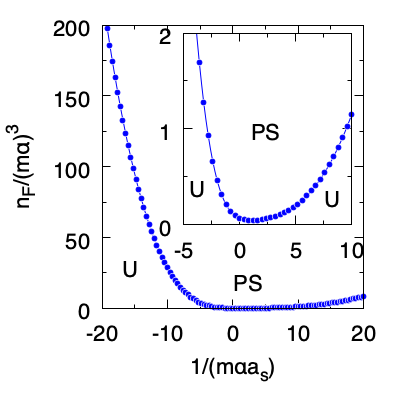}}}
\caption{\label{fig:mb} 
Critical boundary between the uniform superfluid (U) and phase separation 
(PS) that is determined by the effective weakly-interacting Bose-Fermi mixture 
description of a population-imbalanced Fermi gas in the Born approximation.
}
\end{figure}

For instance, it is well-known that a weakly-interacting Bose-Fermi mixture 
is unstable against phase separation with a negative compressibility when 
the density of fermions $n_F$ satisfies~\cite{iskin11}
$
n_F \ge 4\pi^4 U_{BB}^3/(3m_F^3 U_{BF}^6)
$
in three dimensions, where
$
U_{BB} = 4\pi a_{BB}/m_B
$
is the repulsive interaction between bosons, and
$
U_{BF} = 4\pi a_{BF}/m_{BF}
$
is the repulsive interaction between fermions and bosons. Thus, the Bose-Fermi
mixture phase separates when
\begin{align}
\label{eqn:nF}
n_F \ge \frac{4\pi}{3} \frac{m^3m_B^3}{(m_B+m)^6} \frac{a_{BB}^3}{a_{BF}^6},
\end{align}
and is otherwise uniform. By plugging the Born approximations Eqs.~(\ref{eqn:abfborn}) 
and~(\ref{eqn:abbborn}) into Eq.~(\ref{eqn:nF}), we obtain the corresponding relation 
for the stability of a population-imbalanced Fermi gas with SOC. The critical boundary
between the uniform superfluid and phase separation is shown in Fig.~\ref{fig:mb}. 

Here we remark in passing that one can study BCS-BEC evolution for any given 
$a_s$ by tuning the strength $\alpha$ of the SOC, no matter how small or large the 
value of $a_s$ is and independently of its sign. Its physical mechanism is the 
SOC-induced enhancement of $\varepsilon_b$ through the increase of the single 
particle density of states. In particular, when $\alpha$ is large, the nature of the bosons 
that make up the BEC is determined solely by $\alpha a_s$. 
For this reason, these bosons are sometimes called \textit{rashbons} in the recent 
literature since their properties are determined by SOC alone.
See Refs.~\cite{zhai11, iskin11, hu11, he12a, he12b, shenoy12a, shenoy12b} for 
further discussion, including the effective Gross-Pitaevskii description of the 
weakly-interacting dimers in the BEC limit.

\section{Anisotropic (Rashba) spin-orbit coupling}
\label{sec:Rsoc}

Our results can be easily generalized to anisotropic SOC fields. For instance, in the 
presence of a Rashba SOC, the one-body Hamiltonian is governed by
$
h_\mathbf{k} = \varepsilon_\mathbf{k} \sigma_0 + \alpha \mathbf{k_\perp} \cdot \boldsymbol{\sigma},
$
where $\mathbf{k} = (\mathbf{k_\perp}, k_z)$ and $\mathbf{k_\perp} = (k_x, k_y)$, 
leading to
$
\varepsilon_\mathbf{k}^s = (k_\perp^2 + k_z^2)/(2m) + s \alpha k_\perp.
$
Therefore, the ground state of the $-$-helicity band corresponds to a degenerate 
ring of $\mathbf{k_\perp}$ states with the radius $k_m = m\alpha$ at $k_z = 0$ and 
energy $\varepsilon_{\mathbf{k}_m}^- = -m\alpha^2/2$.

In the lowest order in $\mathbf{q} = (\mathbf{q_\perp}, q_z)$ and $q_0$, 
Eq.~(\ref{eqn:Dqfull}) has the generic structure of a simple 
pole~\cite{iskin11, hu11, he12a, he12b, shenoy12a, shenoy12b} 
\begin{align}
\label{eqn:DqR}
D(\mathbf{q}, q_0) = \frac{Z} {q_0 - q_\perp^2/(2m_{B,\perp}) - q_z^2/(2m_{B,z}) + \mu_B + i0^+}
\end{align}
where 
$
Z = 8\pi(|\varepsilon_b| - m\alpha^2)/(m\sqrt{m |\varepsilon_b|})
$
corresponds to the residue of the pole,
$
2m/m_{B,\perp} = (2|\varepsilon_b| - m\alpha^2)/(2|\varepsilon_b|) - 
[(|\varepsilon_b|-m\alpha^2)/(2|\varepsilon_b|)] \log(1 - m\alpha^2/|\varepsilon_b|)
$
and $m_{B,z} = 2m$ determine the anisotropic effective mass of the dimer. 
In addition, $\mu_B = |\varepsilon_b|$ is determined by
$
1 / (m \alpha a_s) = \sqrt{|\varepsilon_b|/(m \alpha^2)} - 
\log[\sqrt{|\varepsilon_b|/(|\varepsilon_b| - m\alpha^2)} 
+ \sqrt{m\alpha^2/(|\varepsilon_b| - m\alpha^2)}].
$
This expression is analytically tractable in three limits~\cite{zhai11, iskin11, hu11}: we find that
(i) $\varepsilon_b = -m\alpha^2 - 4m\alpha^2 e^{2/(m\alpha a_s)-2}$ and $m_{B,\perp} = 4m$ 
in the limit when $1/(m \alpha a_s) \ll -1$,
(ii) $\varepsilon_b \approx -1.44 m\alpha^2$ and $m_{B,\perp} \approx 2.40m$
in the unitarity limit when $1/(m \alpha a_s) = 0$, and 
(iii) $\varepsilon_b = - 1/(ma_s^2)$ and $m_{B,\perp} = 2m$ 
in the limit when $1/(m \alpha a_s) \gg 1$.
Note that the latter limit recovers the usual two-body problem with no SOC 
in the $1/(m \alpha a_s) \gg 1$ limit when $\alpha \to 0^+$.

\begin{figure} [htb]
\centerline{\scalebox{0.43}{\includegraphics{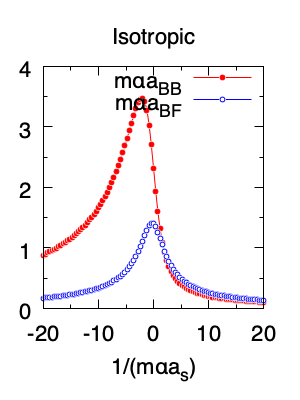} \includegraphics{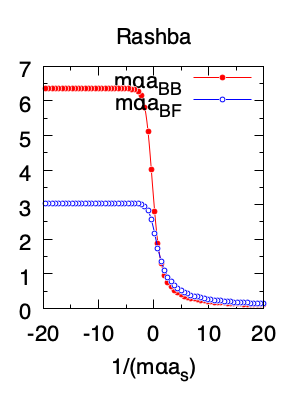}}}
\caption{\label{fig:rashba} 
Dimer-dimer (red) and fermion-dimer (blue) scattering lengths for the isotropic SOC 
(left: same as in Fig.~\ref{fig:abf}) versus Rashba SOC (right) 
in the Born approximations.
}
\end{figure}

Since 
$
k = (m\alpha \widehat{\mathbf{k_\perp}}, -m\alpha^2/2)
$ 
corresponds to the lowest-energy eigenstate in the $-$-helicity band, 
Eq.~(\ref{eqn:tkssb}) gives 
$
{t_k^{--}}^{(0)} (0) = 1/(|\varepsilon_b| - m\alpha^2).
$
By plugging it into Eq.~(\ref{eqn:abft}), we find
\begin{align}
\label{eqn:abfbornR}
a_{BF}^\mathrm{Born} = \frac{2 m_{BF}} {m \sqrt {m |\varepsilon_b|}}
\end{align}
in the Born approximation, where $m_B$ refers to the geometric mean 
$(m_{B,\perp}^2 m_{B,z})^{1/3}$ of the anisotropic effective mass~\cite{shenoy12b}. 
In Fig.~\ref{fig:rashba}, we show $a_{BF}^\mathrm{Born}$ 
as a function of $1/(m \alpha a_s)$, which is analytically tractable in three limits: 
(i) $a_{BF}^\mathrm{Born} \approx 3.04 / (m \alpha)$ in the limit when 
$1/(m \alpha a_s) \ll -1$, 
(ii) $a_{BF}^\mathrm{Born} \approx  2.31 / (m \alpha)$ 
in the unitarity limit when $1/(m \alpha a_s) = 0$, and
(iii) $a_{BF}^\mathrm{Born} = 8a_s/3$ in the limit when $1/(m \alpha a_s) \gg 1$.
Note again the latter limit recovers the usual three-body problem with no SOC 
in the $1/(m \alpha a_s) \gg 1$ limit when $\alpha \to 0^+$.

Similarly, Eq.~(\ref{eqn:tbb0}) gives
$
{t_0^{BB}}^{(1)} (0) = m\sqrt{m}(|\varepsilon_b| + m\alpha^2) / 
[16 \pi (|\varepsilon_b| - m\alpha^2)^2\sqrt{|\varepsilon_b|}],
$ 
and by plugging it in Eq.~(\ref{eqn:abb}), we find
\begin{align}
\label{eqn:abbbornR}
a_{BB}^\mathrm{Born} = \frac{m_B (|\varepsilon_b| + m\alpha^2)} 
{m\sqrt{m} \sqrt{|\varepsilon_b|}^3},
\end{align}
in the Born approximation. In Fig.~\ref{fig:rashba}, we show $a_{BB}^\mathrm{Born}$ 
as a function of $1/(m \alpha a_s)$, which is analytically tractable in three limits: 
(i) $a_{BB}^\mathrm{Born} \approx 6.35 / (m\alpha)$ in the limit when 
$1/(m \alpha a_s) \ll -1$, 
(ii) $a_{BB}^\mathrm{Born} \approx  3.19 / (m \alpha)$ 
in the unitarity limit when $1/(m \alpha a_s) = 0$, and
(iii) $a_{BB}^\mathrm{Born} = 2a_s$ in the limit when $1/(m \alpha a_s) \gg 1$.
Note again that the latter limit recovers the usual four-body problem with no SOC 
in the $1/(m \alpha a_s) \gg 1$ limit when $\alpha \to 0^+$.

In contrast to the isotropic SOC case where $a_{BF}$ and $a_{BB}$ are non-monotonous 
functions of $1/(m \alpha a_s)$, they evolve monotonously in the Rashba SOC. 
Their saturations in the $1/(m \alpha a_s) \ll -1$ limit are caused by the exact cancellation 
of the decay of $Z$ with the divergence of the t-matrices. The decay is faster in the 
isotropic case, causing the peak in the intermediate region. Despite this major difference, 
the isotropic and Rashba SOC cases share some common properties. For instance, the 
decrease, increase and saturation of $a_{BF}$ are in full coordination with those of $a_{BB}$. 
In addition, we note that $a_{BF}$ is greater (smaller) than $a_{BB}$ in approximately 
the $1/(m \alpha a_s) \gtrless 0$ regions.

\section{Conclusion}
\label{sec:conc}

In summary, we studied how SOC affects the fermion-dimer and dimer-dimer 
scattering lengths in the Born approximation, and benchmarked their accuracy 
with the higher-order approximations. We considered both isotropic and Rashba 
couplings in three dimensions, and found that the Born approximation gives 
accurate results for both $a_{BF}$ and $a_{BB}$ in the $1/(m \alpha a_s) \ll -1$ limit.
This is because while the higher-loop contributions form a perturbative series
in the $1/(m \alpha a_s) < 0$ region, they are of similar order in the 
$1/(m \alpha a_s) > 0$ region. We found that the perturbations are controlled 
by the residue $Z$ of the dimer propagator, which decays to $0$ in the 
$1/(m \alpha a_s) \to -\infty$ limit, and increases as $\sqrt{|\varepsilon_b|}$ 
in the $1/(m \alpha a_s) > 0$ region. Therefore, it may be sufficient to consider 
a finite number of higher-order loop diagrams in the $1/(m \alpha a_s) < 0$ region.

On the other hand, a proper description of the $1/(m \alpha a_s) > 0$ region 
requires infinitely-many loop diagrams at all orders. In the case of three-body 
problem, we derived a coupled set of integral equations for the exact atom-dimer 
scattering length, but its numerical solutions remain an open problem. It may be
possible to solve the exact three-body problem through partial-wave expansion,
and address the possibility of a three-body bound state in this system. 
In addition, one may also study the importance of the full momentum and/or full 
frequency dependences of the dimer propagator in the three- and/or four-body 
problems.

\begin{acknowledgments}
The author acknowledges funding from T{\"U}B{\.I}TAK Grant No. 11001-118F359.
\end{acknowledgments}

\appendix

\section{Binding energy and effective mass of the dimer}
\label{sec:app}

\begin{figure} [htb]
\centerline{\scalebox{0.43}{\includegraphics{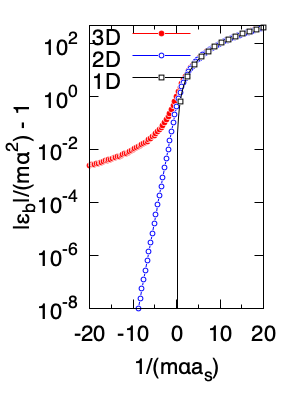} \includegraphics{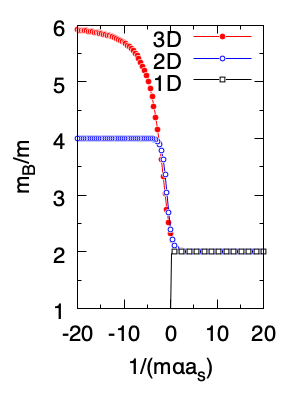}}}
\caption{\label{fig:EbmB} 
Binding energy $|\varepsilon_b| - m\alpha^2$ and effective mass $m_{B,x}$ of the dimer.
Here, 3D SOC field is shown in red and refers to $\alpha \mathbf{k}$ with $\mathbf{k} = (k_x, k_y, k_z)$, 
2D one is shown in blue and refers to $\alpha \mathbf{k_\perp}$ with $\mathbf{k_\perp} = (k_x, k_y, 0)$, 
and 1D one is shown in black and refers to $\mathbf{k} = (k_x, 0, 0)$.
}
\end{figure}

For the sake of completeness, we present the binding energy $|\varepsilon_b| - m\alpha^2$ 
and effective mass $m_B$ of the dimer in Fig.~\ref{fig:EbmB}, where 3D SOC 
field refers to $\alpha \mathbf{k}$ with $\mathbf{k} = (k_x, k_y, k_z)$, 
2D one to $\alpha \mathbf{k_\perp}$ with $\mathbf{k_\perp} = (k_x, k_y, 0)$, 
and 1D one to $\mathbf{k} = (k_x, 0, 0)$. The latter case is trivial because
as the 1D SOC field can be gauged away from the $\mathbf{k}$-space 
integrations, it is equivalent to the usual two-body problem with no SOC~\cite{iskin11}. 
Therefore, a two-body bound state exists only when $a_s > 0$ with an 
effective mass $m_B = 2m$ that is isotropic in space.

In contrast to the 1D case, a two-body bound state exists for all $a_s$ in both 3D 
(isotropic) and 2D (Rashba) SOC fields, which is caused by the increase in the 
low-energy density of one-body states~\cite{zhai11, iskin11, hu11, he12a, he12b, 
shenoy12a, shenoy12b}. In addition, while the effective mass of the 
dimer is isotropic in the 3D case where $m_B = m_{B,x} =  m_{B,y} =  m_{B,z}$ is 
shown in the figure, it is anisotropic in the 2D case where only 
$m_{B,\perp} = m_{B,x} = m_{B,y}$ is shown in the figure, and $m_{B,z} = 2m$ 
for all $a_s$.


\begin{thebibliography}{99}
\bibitem{integralnote} 
To achieve a reliable numerical convergence, we change the limits of the 
radial $dq$ integration from $[0, \infty]$ to $[1, -1]$ via the change 
of variables $q' = 2/(q + 1) - 1$, i.e., 
$
\int_0^\infty f(x)dx = 2 \int_{-1}^{1} f[2/(x+1)-1] dx/(x+1)^2
$
\end{thebibliography}
\end{document}